\documentclass[12pt,a4paper]{article}

\usepackage{feynmp,amsbsy,latexsym,cite}

\newcommand{\no}{\noindent}

\newcommand{\Ha}{{\cal H}}

\newcommand{\mod}[1]{\vert {#1}\vert}
\newcommand{\pa}{\partial}

\newcommand{\ee}{\mathrm{e}}

\newcommand{\psidirac}{\psi_D}

\newcommand{\xb}{{\boldsymbol x}}
\newcommand{\yb}{{\boldsymbol y}}
\newcommand{\zb}{{\boldsymbol z}}

\newcommand{\PP}{\mathsf{P}}

\begin{document}

\begin{titlepage}

\rightline{PLY-MS-99-91}

\vskip19truemm
\begin{center}{\Large{\textbf{Hadrons Without Strings}}}\\ [12truemm]
\textsc{Martin Lavelle}\footnote{email: mlavelle@plymouth.ac.uk} and
\textsc{David McMullan}\footnote{email: dmcmullan@plymouth.ac.uk}\\
[5truemm] \textit{School of Mathematics and Statistics\\ The
University of Plymouth\\ Plymouth, PL4 8AA\\ UK} \end{center}

\bigskip\bigskip\bigskip
\begin{quote}
\textbf{Abstract:} Descriptions of hadrons and glueballs
can be constructed using strings to preserve gauge invariance. We show
how this string dependence may be removed at all orders in
perturbation theory.
\end{quote}

\end{titlepage}

\setlength{\parskip}{1.5ex plus 0.5ex minus 0.5ex}

\noindent \no \textbf{Introduction}

\smallskip

\no Gauge invariant descriptions of hadrons and their substructures
are needed to directly link phenomenology to QCD.
The need for gauge invariance becomes clear when one thinks of jet
formation: here a quark and an anti-quark move away from each
other and turn into two independent jets of hadrons. To
obtain such colourless objects, without making the quark matter directly interact
so that jet formation becomes implausible, one needs to surround the quarks
with glue, which makes the system locally gauge invariant. It is then
energetically favoured for
this glue to split into two as the quark matter
fields reach a large enough separation. This is naively described by parton
fragmentation~\cite{Field:1978fa}
or, closer to QCD, modelled by a gluonic string snapping~\cite{Andersson:1983ia}. In reality, of course, the
glue will have a much richer structure.

It is known~\cite{Lavelle:1996tz} that colour charge is \emph{only} defined on locally
gauge invariant states and thus all physically meaningful hadronic
substructures  -- coloured quarks in jet formation, constituent
quarks, the form of the glue in glueballs and the pomeron -- must
be described in a gauge invariant way. The appropriate variables
for short and long distance physics will vary, however, there are
far too few methods known with which they can be described. In this
paper we want in particular to show how to proceed from
string-based to constituent-type descriptions. We will argue that
these latter variables provide a better starting point for probing
the above structures.

\begin{figure}[!hbp]
\begin{center}
\includegraphics{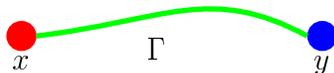}
\end{center}
\caption{Using strings to try to describe $e^+e^-$ states.}
        \label{meson}
\end{figure}

Non-local states and currents which are
supposed to overlap with hadrons are generally made gauge invariant
with the insertion of \lq strings\rq,
i.e., the fundamental fields are linked by path ordered
exponentials. To illustrate this, consider the case of Quantum Electrodynamics (QED)
where we can make a candidate description of positronium by
attaching a string that runs from one fermion to the other over a path $\Gamma$ as depicted in
Fig.~\ref{meson}.

This system (which we for simplicity keep in a single time slice)
\begin{equation}\label{deco}
\bar\psi(t,\xb) \exp\left\{
  -i e \int_\Gamma dz_i A_i(t,\zb)
  \right\}
  \psi(t,\yb)\,,
\end{equation}
is gauge-invariant, however, it
has a manifest path-dependence which has no obvious physical
interpretation. It is, though, possible to remove this path
dependence using the standard decomposition of $A_i$ into
longitudinal $(A_i^L)$ and gauge-invariant, transverse ($A_i^T$)
degrees of freedom: $A_i=A_i^L+A_i^T$ where $A_i^L=\partial_i
\partial_j A_j/\nabla^2$. We so obtain
\begin{equation}\label{two}
\bar \psi(x)\ee^{  i e \frac{\partial_i A_i}{\nabla^2}(x)}
\ee^{-  i e \int_\Gamma dz_i A_i^T(t,\zb)
}
\ee^{  -i e \frac{\partial_i A_i}{\nabla^2}(y)
}
\psi(y)\,,
\end{equation}
where $\nabla^{-2}$ is the standard Green's function for the spatial
Laplacian.

Note that by writing the positronium state in this way we have put
all of the arbitrary $\Gamma$-dependence into a separately gauge
invariant factor and it is natural to
drop it. The remaining gauge-invariant combination
\begin{equation}\label{psiD}
 \bar \psidirac(x)\psidirac(y):=
 \bar \psi(x)\ee^{  i e \frac{\partial_i A_i}{\nabla^2}(x)}\ee^{ - i e \frac{\partial_i A_i}{\nabla^2}(y)
}
\psi(y)\,,
\end{equation}
is a product of two individually gauge invariant terms, $\psidirac$, which were
first written down by Dirac~\cite{Dirac:1955ca} and have been employed by a
variety of authors in the intervening period (for a review,
see~\cite{Lavelle:1997ty}).

Further strong physical reasons for dropping
the path dependent factor in (\ref{two}) can be seen by considering
the energy of the (straight line) string state.
To find the expectation value of the Hamiltonian,
$\Ha=\frac12\int d^3w (E_i^2(w)+B_i^2(w))$,
for such states, we employ the standard
equal-time commutators, $[E_i(\xb),A_j(\yb)]=i\delta_{ij}
\delta(\xb-\yb)$. The magnetic field commutes with (\ref{meson})
but the electric field does not. We straightforwardly
find for the potential energy of
this state~\cite{Haagensen:1997pi}
\begin{equation}\label{confines!}
  V(r) =
\frac{e^2}2\delta^{(2)}_\perp(0) r
\,,
\end{equation}
where $r=\mod{\xb-\yb}$
and we have dropped the divergent self-energy contributions (the $r$ independent
terms). This shows a linear, confining potential
with a divergent coefficient -- in QED! It is clear that \lq
confinement\rq\ here is an artefact
due to trapping the electric flux in a string and that the divergence of the
coefficient is a result of the string being infinitely thin (the $\perp$
subscript indicates that this divergent factor originates from the directions
orthogonal to the path of the string).

Of course this, although gauge invariant, is not physical. It
corresponds to an infinitely excited state and is not
stable~\cite{Prokhorov:1993}. A
discussion of this instability and an animation of the decay of such
a state can be found at \texttt{http://www.ifae.es/\~{}roy/qed.html}.
We can gain further insight from recalling that Gauss' law tells us
that $\pa_i E_i= e\rho$, where $\rho$ is the charge density. If we
now decompose the electric field into transverse and longitudinal
components, the contribution of the electric field to the above
Hamiltonian can be so re-expressed as a sum of two positive
semi-definite terms, $\frac12\int d^3w ({E_i^T}^2+{E_i^L}^2)$. The
integral over the longitudinal part of the electric field is
evidently fixed by Gauss' law to be the same for \emph{any} gauge
invariant description of a state with a fixed number of charges.
However, the transverse component, $E_i^T$, is not so fixed. For the
above \lq stringy\rq\ state it in fact diverges, while for the
variable $\psi_D$ it is zero which minimises the energy. Hence the
stringy state is unstable and it decays into a product of two charged
fields as given in~(\ref{psiD}). It is easy to repeat the analysis
leading to (\ref{confines!}) for these constituents and so see that
the familiar potential of electrodynamics, $-e^2/4\pi r$, holds for
this state. This implies that in any string description of this
$e^+e^-$~state, where we can factor out the exponential over the
$A_i^T$~variables, we indeed ought to drop this factor and work with
a product of two $\psi_D$ variables.

The string-based description has, we have seen, many unattractive
characteristics which make it a poor starting point to analyse
positronium. The alternative, constituent picture has a much more
appealing physical base. Its domain of validity is for heavy charges
since the charges, $\psidirac$, have to be
static~\cite{Lavelle:1997ty}. To approach a more realistic
description, one would need to expand in the masses and so build in
binding energies. However, since this starting point is energetically
favoured and has a clear physical interpretation it is
a more sensible starting point for studying bound states in QED.

The question now arises, can a similar decomposition be carried
out for QCD?  The string picture, since the matter fields would
feel confining interaction at short distances, does not include the short distance
forces felt by quarks and gluons. However, the generalisation of $\psi_D$
would include the Coulombic type of interaction which dominates at short
distances. It is not, however, initially obvious that this can be done.
The vector boson fields are now Lie algebra valued and any
$T^a$~matrices sandwiched between spinors cannot be extracted.
Furthermore the decomposition into longitudinal and
gauge-invariant, transverse degrees of freedom used above does not
hold in QCD. In the rest of this paper we will show that these
obstacles can in fact be overcome at any order in perturbation theory
and the path dependence can
be removed in QCD. In the mesonic case we will  thus
regain the constituent structure which has been previously shown~\cite{Lavelle:1998dv} by us to
be responsible for asymptotic freedom.

\bigskip

\no \textbf{Non-Abelian Strings}

\smallskip

\no We now want to recall  how to
introduce into a non-abelian theory variables which are both gauge invariant
and do not have a path dependence. Then we will show the relation between these
degrees of freedom and the string-based systems at lowest order in
the coupling.

We define
$h^{-1}$ to be a field dependent element of the gauge group that
transforms as
\begin{equation}\label{htrans}
  h^{-1}(x)\to  h^{-1}(x) U(x)\,.
\end{equation}
Recalling how  fermions transform under gauge transformations,
$\psi\to U^{-1}\psi$, it is clear that the following fields
\begin{equation}\label{psih}
 \psi_h(x)= h^{-1}\psi(x)\,,
\end{equation}
describe
gauge-invariant quarks.
This is the non-abelian analogue of $\psidirac$. We call $h^{-1}$ a
\textsl{dressing}, it corresponds here to the gluonic cloud which
surrounds a static fermion.

To make this explicit, we now recall that there is a simple
generalisation of (\ref{psiD}) to all orders in perturbation
theory (see the appendix of~\cite{Lavelle:1997ty} for a
discussion of how to generate it).
Essentially we have
\begin{equation}\label{hdef}
  h^{-1}=\exp\{  (g\chi^a_1+g^2\chi^a_2+g^3\chi^a_3+\cdots)T^a \}\,,
\end{equation}
where
\begin{equation}\label{chidef}
\chi_1^a=\frac{\partial_j A_j^a}{\nabla^2}\,,\quad
   \qquad
\chi_2^a=f^{abc}\frac{\partial_j}{\nabla^2}\left(
\chi_1^bA_j^c+\frac12(\partial_j\chi_1^b)\chi^c_1
\right)
  \,,\qquad \mbox{etc}\,.
\end{equation}
Here $\chi_1^a$ is, up to a trivial colour factor,
what we had above in QED while $\chi_2^a$~etc
are minimal extensions whose role is to preserve the
gauge transformation property, Eq.~\ref{htrans}.

In a previous letter~\cite{Lavelle:1998dv} we
have demonstrated that precisely this gluonic configuration
in $h^{-1}$ is responsible for antiscreening. We now want to show how we
can use these descriptions to remove the string
dependence in the QCD~equivalent of Fig.~\ref{meson}:
\begin{equation}\label{mes}
\bar\psi(x)\PP \exp\left(g\int^y_xdz_iA_i(z)
  \right)\psi(y)\,,
\end{equation}
which is gauge invariant
but path dependent.
At lowest order in perturbation
theory in QCD we can make the same decomposition of the vector
potential, since $A_i^T$ is gauge invariant up to terms of order
$g$. We can thus rewrite (\ref{mes}) as
\begin{equation}
\bar\psi(x)h(x)\PP \exp\left(g\int^y_xdz_iA_i^T(z)
  \right)h^{-1}(y)\psi(y)+O(g^2)\,,
\end{equation}
where $h^{-1}$ is the static dressing. We cannot now factor off
the exponential due to the presence of the Gell-Mann matrices. However, if we
write out the last equation more explicitly:
\begin{equation}\label{mss}
\bar\psi(x)\left(1-g\frac{\partial_iA_i^aT^a}{\nabla^2}(x)\right)
\left[1+g\int^y_xdz_iA_i^T(z)\right]
\left(1+g\frac{\partial_iA_i^aT^a}{\nabla^2}(y)\right)
\psi(y) +{{O}}(g^2)
\,,
\end{equation}
then, since at this order the terms in the square bracket are gauge
invariant, we may discard the path dependent term,
keeping only the 1, to obtain,
at this lowest order, a result equivalent to that which we would
obtain from multiplying two dressed quarks, $\psi_h$, together. Eq.~\ref{mss}
is suggestive of how this will generalise to higher orders. We
stress again that the term which we drop has a singular energy
since the flux is concentrated onto a line.

\medskip

Before moving on to the general
proof, let us consider the baryonic case at $O(g)$.
\begin{figure}[tbp]
\begin{center}
%
\includegraphics{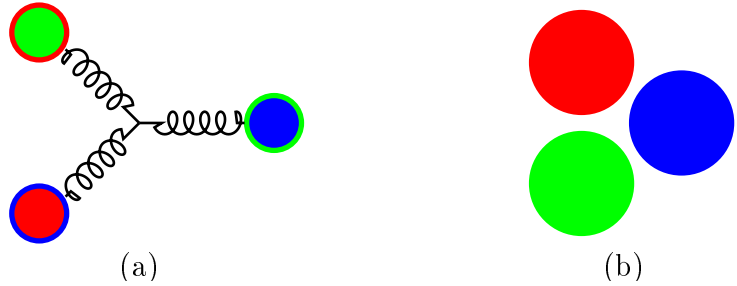}
\end{center}
\caption{Two descriptions of a baryon: \emph{a)} using strings; \emph{b)} in terms
of constituents. Note that \emph{a)} also
possesses a dependence on the point where the strings meet and that only in \emph{b)}
are there locally gauge invariant substructures with well-defined colour.}
        \label{baryons}
\end{figure}
Essentially we want to show that the structure of
Fig.~{\ref{baryons}}\,a can be rewritten in perturbation
theory as a product of three gauge invariant quarks, each then
with well-defined colour, plus extra additive
terms which contain the (singular) path dependence and should be dropped.

The fermions in local baryonic currents\footnote{The exact Lorentz structure of the
current is unimportant.} such as $\epsilon^{abc}\psi^a\psi^b\psi^c$ may be
taken to different positions in a gauge invariant way by attaching
strings as follows
\begin{eqnarray}\label{brr}
&  &\left[\PP \exp\left(g\int^x_w dz_iA_i(z)
  \right)\psi(x)\right]^a
\left[\PP \exp\left(g\int^y_w dz_iA_i(z)
  \right)\psi(y) \right]^b \cr
& &\qquad \qquad\qquad \times
\left[\PP \exp\left(g\int^l_w dz_iA_i(z)
\right)\psi(l) \right]^c\epsilon^{abc}
\,.
\end{eqnarray}
If we decompose the vector potentials once more into transverse and
longitudinal components we obtain at $O(g)$
\begin{eqnarray}\label{hwtodie}
&  &\left[h(w)\left(1+g\int^x_w dz_iA_i^T(z)
  \right)h^{-1}(x)\psi(x)\right]^a\cr
  && \times
\left[h(w)\left(1+g\int^y_w d{z'}_jA_j^T(z')
  \right)h^{-1}(y)\psi(y)\right]^b\cr &&{}\phantom{\times}\times
\left[h(w)\left(1+g\int^l_w d{z''}_kA_k^T(z'')
  \right)h^{-1}(l)\psi(l)\right]^c+O(g^2)
\epsilon^{abc}
\,.
\end{eqnarray}
Again discarding the singular path dependent terms, we obtain at this order
two types of terms: those corresponding to an (antisymmetric)
product of three dressed quarks at $x$, $y$ and $l$ and also terms
of the form
\begin{equation}
g\psi^a(x)\psi^b(y)\left[ \frac{\partial_iA_i^T(w)}{\nabla^2}
\psi(l)\right]^c\epsilon^{abc} +\hbox{2 similar}
\,,
\end{equation}
which come from expanding the $h(w)$ factors in~(\ref{hwtodie}).
These terms are not path dependent, but they appear to have an
unattractive dependence upon the arbitrary point, $w$, where the strings join.
However, we can show that they vanish; to do this rename the dummy
indices to extract an overall factor of
\begin{equation}
(T^\alpha)_{cd}\epsilon^{abc} +
(T^\alpha)_{cb}\epsilon^{acd} +
(T^\alpha)_{ca}\epsilon^{cbd}
\,,
\end{equation}
which vanishes due to the tracelessness of the Gell-Mann matrices.
We thus see that on extracting and discarding the path dependence we
may,  at lowest order in the coupling,
write (\ref{brr}) as a product of three
dressed quarks, $\psi_h$, each of which has a well-defined colour.
This system then depends neither on the arbitrary paths nor on their junction
point.

It should be clear that we could also
attach a string to an individual quark or
connect field strengths to produce non-local
descriptions of glue. We will return to these applications below.
Now though we wish to show how these arguments can be extended to
all orders.

\bigskip

\no \textbf{All Orders Strings}

\smallskip

\no  We now want to show how to express path ordered exponentials over the vector
potentials of QCD in terms
of products of path ordered exponentials over the gauge invariant gluons and
the dressings. This will permit us to remove the contour
dependence to any order of perturbation theory.

Recall that a path ordered exponential~\cite{Feynman:1951gn}
transforms under a gauge transformation as
\begin{equation}\label{pdef}
  \PP \exp\left(g\int^y_xdz_iA_i(z)
  \right)\to
U^{-1}(x)\PP \exp\left(g\int^y_xdz_iA_i(z)
  \right)U(y)\,,
\end{equation}
and, since a fermion transforms as $\psi\to U^{-1}\psi$, we can
produce gauge invariant composites by linking quarks together via
such strings.

We now want to demonstrate
that we can rewrite this at all orders as
\begin{equation}\label{relate}
  \PP \exp\left(g\int^y_xdz_iA_i(z)
  \right)=
 h(x) \PP \exp\left(g\int^y_xdz_iA_i^h(z)\right) h^{-1}(y)
\,,
\end{equation}
where the external dressing factors are gauge dependent and the
central term is a path ordered exponential over the gauge invariant
field, $A^h$. There are no ordering
problems for the glue dressed in this fashion.

It is also clear that (\ref{relate}) transforms as one would expect
of a path ordered exponential:
\begin{equation}\label{transf}
 h(x)
 \PP \exp\left(g\int^y_xdz_iA_i^h(z)\right) h^{-1}(y)
 \to
U^{-1}(x)
h(x) \PP \exp\left(g\int^y_xdz_iA_i^h(z)\right) h^{-1}(y)U(y)
\,.
\end{equation}

To prove (\ref{relate}), recall the formal definition
of a path ordered exponential~\cite{Feynman:1951gn}
\begin{equation}\label{deff}
  \PP \exp\left(g\int^y_xdz_iA_i(z)
  \right)\!=\!\lim_{N\to\infty}\prod^{N-1}_{0}
  \left[1+gA_i(z(k))\left\{z_i(k+1)-z_i(k)\right\} \right]
\,,
\end{equation}
where the end points $z(0)$ and $z(N)$ are fixed at $x$
and $y$ respectively.

If we consider one element of this product, which we write as
$1+gA_i(w)\{w_i-s_i\}$, and take $\{w_i-s_i\}$ to be small then we want to
show that this term can be written using the dressing as
\begin{equation}\label{lnk}
1+gA_i(w)\{w_i-s_i\}=h(s)\left[
1+gA_i^h(w)\{w_i-s_i\}\right]h^{-1}(w)
\,.
\end{equation}
If we can show this for one term, then (\ref{relate}) follows by
inspection. We write the 1 as $h(s)h^{-1}(s)$ and Taylor expand
$h^{-1}(s)$ around $w$. Dropping terms of order $\{w_i-s_i\}^2$ we
so obtain
\begin{equation}
h(s)\left(1+g\left( h^{-1}(s)A_i(s)h(s)
-\frac1g (\partial_i h^{-1}(s)) h(s)
\right)\{w_i-s_i\}\right)h^{-1}(w)
\,,
\end{equation}
and using $(\partial_i h^{-1}(s)) h(s)=- h^{-1}(s)
\partial_ih(s)$, we obtain (\ref{lnk}) and
have thus demonstrated (\ref{relate}).

As well as this formal all orders proof, we have explicitly
checked (\ref{relate}) at higher orders.
With this result we will now find it relatively straightforward
to extract the
string dependence from descriptions of hadrons in a gauge
invariant way.

\bigskip
\no \textbf{Making Hadrons}

\smallskip

\noindent Let us now look at the following in turn:
quarks, mesons, baryons and finally glue.

\noindent a) A \textit{quark} with
a string attached to it may be written
as
\begin{equation}
\PP \exp\left(g\int^y_\infty dz_iA_i(z)
  \right)\psi(y)=
h(\infty) \PP
 \exp\left(g\int^y_\infty dz_iA_i^h(z)\right) h^{-1}(y)\psi(y)
\,,
\end{equation}
where the string is taken along some path out to infinity. Since
for reasons of finite energy we require that our fields vanish at
spatial infinity, we have $h(\infty) \to 1$ and we can also
factor out the gauge invariant (but contour dependent) path ordered
exponential, leaving ourselves with  a dressed quark,
$h^{-1}(y)\psi(y)$. This shows how one can proceed from the gauge invariant
variables introduced by Mandelstam~\cite{Mandelstam:1968hz} to those of Dirac.

\noindent b) A \textit{mesonic} structure may be made up by
linking two fermions by a string
\begin{equation}
\bar\psi(x)\PP \exp\left(g\int^y_x dz_iA_i(z)
  \right)\psi(y)
= \bar\psi(x)h(x) \PP
 \exp\left(g\int^y_x dz_iA_i^h(z)\right) h^{-1}(y)\psi(y)
\,,
\end{equation}
where we again have used (\ref{relate}). If we expand the path
ordered exponential in $g$ then only the lowest term, unity, is
independent of the string's path. Since by construction the higher
order terms are gauge invariant, we can drop them and so obtain
the non-local, gauge invariant and path independent system,
$\bar\psi(x)h(x) h^{-1}(y)\psi(y)$. Note that we here cannot
factor out the $\Gamma$~dependence but are free to
drop this additive, singular term.

\noindent c) For \textit{baryonic} structures we recall (\ref{brr})
which, using our result (\ref{relate}) and  discarding as in the
mesonic case
all the path dependent terms, can be expressed as
\begin{equation}
\left[h(w)h^{-1}(x)\psi(x)\right]^a
\left[h(w)h^{-1}(y)\psi(y) \right]^b
\left[h(w)h^{-1}(z)\psi(l) \right]^c\epsilon^{abc}
\,,
\end{equation}
which can be rewritten as
\begin{equation}\label{bryn}
h(w)^{aa'} h(w)^{bb'} h(w)^{cc'}
\left[\psi^h(x) \right]^{a'}
\left[\psi^h(y) \right]^{b'}
\left[\psi^h(l) \right]^{c'}\epsilon^{abc}
\,.
\end{equation}
Since both the entire expression  and
the individual dressed quarks are by
construction gauge invariant, $h(w)^{aa'} h(w)^{bb'}
h(w)^{cc'}\epsilon^{abc}$ must also be gauge invariant. However,
we can always choose a gauge such that $h(w)=1$, and so
$h(w)^{aa'} h(w)^{bb'}h(w)^{cc'}\epsilon^{abc}= \delta^{aa'}
\delta^{bb'}\delta^{cc'}\epsilon^{abc}$ in all gauges.
Therefore (\ref{bryn})
can be written solely in terms of dressed quarks as
$\psi_a^h(x)\psi_b^h(y)\psi_c^h(l) \epsilon^{abc}$.

d) As far as \textit{glue} is concerned we can similarly link
field strengths with path ordered exponentials and take traces.
The path dependence can then be extracted just as above. Of course we
also have for glue the possibility to just directly combine dressed vector
potentials, $A^h_\mu {A^h}^\mu$. Such gauge invariant structures have not
previously been studied. Condensates of such structures,
non-local, gauge invariant and of dimension 2 could be
phenomenological significant; we note that many authors have
argued for the importance of $\langle A_\mu^2\rangle$
condensates  -- it seems that the stigma of
gauge dependence can be removed from such objects (but see below).

\bigskip

\no \textbf{Conclusions}

\smallskip

\no We have seen in QED that string-like descriptions are
singular and that the path dependence can be removed,
in a gauge invariant way, to
yield the correct lowest energy state for static charges.
In QCD we have demonstrated that such path dependence
can be removed at all orders of perturbation theory. This opens up
new, gauge invariant ways to describe hadrons and their substructure.

\begin{figure}[tbp]
\begin{center}
\includegraphics{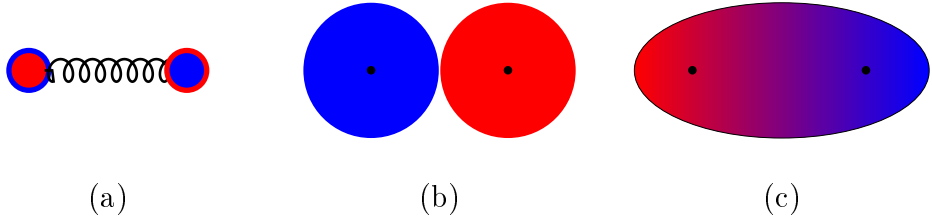}
 \end{center}
\caption{Three different descriptions of a meson. a) Strings: suitable at strong coupling; b) gauge invariant
constituents: describes the short distance glue around current
quarks and appropriate before confinement dominates; c) hadronic variables: correct at larger separations.}
        \label{mesonthree}
\end{figure}

Fig.~\ref{mesonthree} shows three possible ways of describing a
meson in a gauge invariant fashion.
The approach presented in this paper is closest
to Fig.~\ref{mesonthree}b which should describe the physics prior to the onset
of confinement. (Note though that the fall-off of our variables is such that they
extend out to spatial infinity.) In particular they include the
Coulombic-type interaction between quarks.
It is thus evident that,
at short distances, where the forces are essentially
Coulombic, the descriptions we suggest are to be preferred. We note
that these fields have been shown to describe the glue responsible
for anti-screening~\cite{Lavelle:1998dv}. It can be shown that the
constituent picture presented here does not yet generate three-body
forces in baryons. Such forces may emerge from hybrid pictures involving
diquarks.

Infinitesimally thin string-type descriptions are only intuitively
expected to be relevant in the strong
coupling limit, where the shortest path is to be used.
(We note that in weak coupling lattice simulations it is
found~\cite{Cella:1994nn} in
practice that \lq smearing\rq\ the string paths in Wilson loop
calculations helps in numerical calculations.)

Of course, a physical meson, while certainly not well-described by a
string, is not just two constituents and the favoured
glue will be in some sense cigar shaped. The emergence of
such a picture needs to be further pursued; higher order
perturbative studies will be useful in this regard, but
non-perturbative work will be crucial. We note here previous
attempts to construct, in simpler systems, gluonic
distributions which yield a phenomenological potential, of
$r+1/r$~type~\cite{Zarembo:1998ms}.

There are many questions which immediately cry out to be
addressed. The construction presented above holds to all orders
of perturbation theory, yet we know that constituent variables only
have a limited phenomenological validity.
One of the attractive features of our approach is that,
although some non-perturbative effects
can be incorporated into the construction of the dressing, the
construction of  $h^{-1}$ is known to break down in the non-perturbative
domain~\cite{Lavelle:1997ty}. Mapping out its domain of validity is a
major focus of our programme. An immediate corollary is that the $\langle A^hA^h
\rangle$ condensate can only have a limited validity. The physical
implications of this fact need to be studied.

 We also note that path-ordered
exponentials are widely used in the literature to create non-local,
gauge invariant states. We believe that the path independent approach presented
here offers an opportunity to test the formalism-dependence of the conclusions of such
analyses. From what we have seen above, it seems
natural to use these variables as input for
variational methods to find the
energetically favoured glue in hadronic physics.
Finally, the variables used here describe the glue close to
current quarks and are suitable for short distance
hadronic physics. Parton-hadron duality should we feel be couched in
these terms. As such the route from partons to jets will be modified
by including dressings and thus these methods have potential
implications for jet formation.

\no\textbf{Acknowledgements:} We thank Emili Bagan, Robin Horan and Shogo Tanimura
for discussions.


\end{document}